\documentclass[%
 aip,
 amsmath,amssymb,
 reprint,%
]{revtex4-1}

\usepackage{graphicx}
\usepackage{dcolumn}
\usepackage{bm}

\usepackage[utf8]{inputenc}
\usepackage[T1]{fontenc}
\usepackage{mathptmx}
\usepackage{etoolbox}
\usepackage{color}

\usepackage{comment}

\makeatletter
\def\@email#1#2{%
 \endgroup
 \patchcmd{\titleblock@produce}
  {\frontmatter@RRAPformat}
  {\frontmatter@RRAPformat{\produce@RRAP{*#1\href{mailto:#2}{#2}}}\frontmatter@RRAPformat}
  {}{}
}%
\makeatother
\begin{document}

\preprint{AIP/123-QED}

\title{Performance enhancement of a spin-wave-based reservoir
computing system utilizing different physical conditions}

\author{Ryosho Nakane}
\affiliation{Department of Electronic Engineering and Information Systems, Graduate School of Engineering, \\ The University of Tokyo, 7-3-1 Hongo Bunkyo-ku, Tokyo 113-8656, Japan \\}
\author{Akira Hirose}
\affiliation{Department of Electronic Engineering and Information Systems, Graduate School of Engineering, \\ The University of Tokyo, 7-3-1 Hongo Bunkyo-ku, Tokyo 113-8656, Japan \\}
\author{Gouhei Tanaka}
\affiliation{Department of Electronic Engineering and Information Systems, Graduate School of Engineering, \\ The University of Tokyo, 7-3-1 Hongo Bunkyo-ku, Tokyo 113-8656, Japan \\}
\affiliation{%
International Research Center for Neurointelligence (IRCN), \\ The University of Tokyo, 7-3-1 Hongo Bunkyo-ku, Tokyo 113-0033, Japan \\}
\affiliation{%
Department of Mathematical Informatics, Graduate School of Information Technology
and Science,\\ The University of Tokyo, Tokyo 113‑8656, Japan\\}

\date{\today}

\begin{abstract}
The authors have numerically studied how to enhance reservoir computing performance by thoroughly extracting their spin-wave device potential for higher-dimensional information generation. The reservoir device has a 1-input exciter and 120-output detectors on the top of a continuous magnetic garnet film for spin-wave transmission. For various nonlinear and fading-memory dynamic phenomena distributing in the film space, small in-plane magnetic fields were used to prepare stripe domain structures and various damping constants at the film sides and bottom were explored. The ferromagnetic resonant frequency and relaxation time of spin precession clearly characterized the change in spin dynamics with the magnetic field and damping constant.
The common input signal for reservoir computing was a 1 GHz cosine wave with random 6-valued amplitude modulation.  A basic 120-dimensional  reservoir output vector was obtained from time-series signals at the 120 output detectors under each of the three magnetic field conditions. Then, 240- and 360-dimensional reservoir output vectors were also constructed by concatenating two and three basic ones, respectively. In nonlinear autoregressive moving average (NARMA) prediction tasks, the computational performance was enhanced as the dimension of the reservoir output vector becomes higher and a significantly low prediction error was achieved for the 10th-order NARMA using the 360-dimensional vector and optimum damping constant. 
The results are clear evidence that the collection of diverse output signals efficiently increases the dimensionality effective for reservoir computing, i.e., reservoir-state richness. This paper demonstrates that performance enhancement through various configuration settings is a practical approach for on-chip reservoir computing devices with small numbers of real output nodes.
\end{abstract}

\maketitle

\section{Introduction}
Reservoir computing (RC) is an emerging machine-learning scheme suitable for information processing of time-series data \cite{jaeger2001echo, maass2002real}.  Recently, RC utilizing a nonlinear physical system, called physical RC \cite{tanaka2019recent, nakajima2020physical}, has received much attention from electronics, because its low training cost and high adaptability for various information processing allow us to implement low-power and real-time computing hardware needed for edge computing on the internet of things (IoT) era \cite{abbas2017mobile, shi2016edge}.  So far, physical RC systems with various on-chip reservoir devices have been widely studied, which include device proposals, numerical experiments, and experimental demonstrations \cite{tanaka2019recent, nakajima2020physical, appeltant2011information, torrejon2017neuromorphic, nakane2018reservoir, moon2019temporal, toprasertpong2022, kanao2019reservoir, tanaka2022simulation}.  One of the important requirements for high-performance RC is to extract high-dimensional output node states through rich physical dynamics in response to a time-series input signal, i.e., to obtain vast numbers of diverse temporal output signals from a reservoir device.  In this respect, spin waves are very promising, since nonlinear phenomena, such as nonlinear propagation and interference, can induce space-distributing nonlinear dynamics in a continuous magnetic film \cite{gurevich2020magnetization, stancil2009spin, mahmoud2020introduction,papp2021nanoscale}, which can also exclude an intractable technical problem in fabricating enormous internal wirings.  Furthermore, even after the fabrication of an on-chip device, physical dynamics can be reproducibly changed with the condition of the transmission medium, i.e., the magnetic material, through sensitive adjustment in configuration settings, such as the magnitude of an external magnetic field. 

We proposed a spin-wave-based RC device on a chip and have demonstrated its basic capabilities for information processing, using numerical experiments \cite{nakane2018reservoir, ichimura2021numerical, nakane2021spin}.  
More-recent experimental RC with spin waves also indicates high potentials of this methodology in actual device operations \cite{watt2021implementing, nishioka2022edge}.
At the present stage, high dimensionality in output node states is a key to further boost the computational performance.  Thus, a practical approach for this purpose is strongly needed, where the most desirable strategy is to realize diverse dynamics using a single device structure.

In this study, we develop an efficient method to realize higher dimensionality in output node states for RC, by collecting output signals in response to a time-series input signal while the physical condition of the device is changed by the magnitude of an external magnetic field.  



\begin{figure*}[tb]
\includegraphics[width=\linewidth]{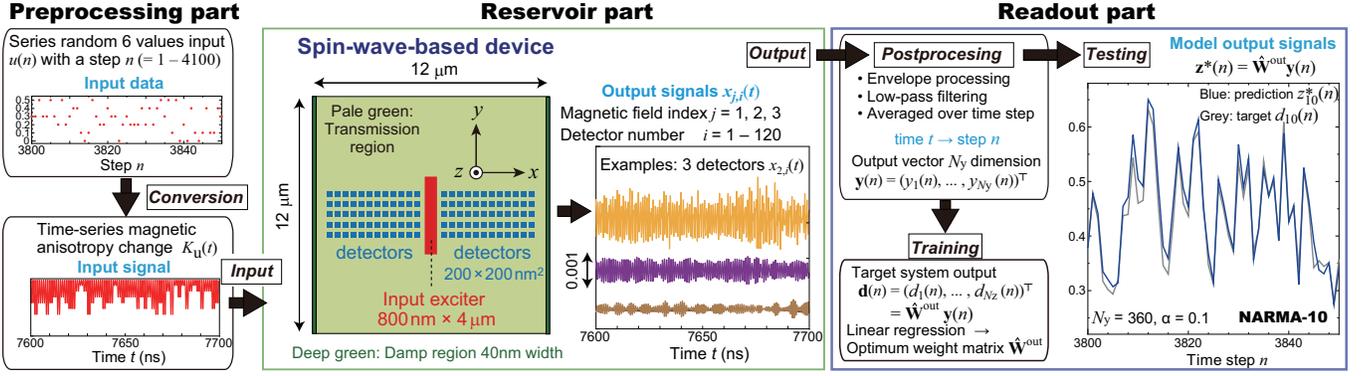}
\caption{\label{FIG_1} Spin-wave-based reservoir computing (RC) system that consists of a preprocessing part, a reservoir part with a spin-wave-based device, and a readout part with adjustable weights. The spin-wave-based device has an input exciter with 800 nm $\times$ 4 $\mu$m $\times$ 40 nm and 120 detectors with 200 nm $\times$ 200 nm $\times$ 40 nm, where the input exciter is placed at the center of the $xy$ plane and 12 $\times$ 5 detectors are regularly arranged in the right and left areas of the $xy$ plane.  In the reservoir part, three output signals in response to the input signal are shown by orange, purple, and brown curves, for which an external magnetic field $\mu_0H^\textup{EX}$ is 0.5 mT ($j$ index is 2) and the damping constant $\alpha$ in the damping region is 0.1. In the readout part, grey and blue curves show the 10th-order nonlinear autoregressive moving average (NARMA-10) target $d_{10}(n)$ and prediction $z^*_{10}(n)$ signals, respectively, in the testing phase of RC.}
\vspace{-10pt}
\end{figure*}


\section{Simulation method, material parameters, and device structure}
Figure \ref{FIG_1} shows our RC system that is composed of a preprocessing part, a reservoir part, and a readout part.  The role of the reservoir part is to nonlinearly transform a time-series input signal to higher-dimensional spatiotemporal output signals.  First, we explain the spin wave device shown as a top view, where  a 320-nm-thick $\textup{Y}_3\textup{Fe}_5\textup{O}_{12}$ (YIG) film with $12 \times 12 \ \mu \textup{m}^2$ in area is used for the spin-wave signal transmission, a Cartesian coordinate system is defined with the origin at the center of the surface plane, one input exciter with 800 nm $\times$ 4 $\mu$m in area and 40 nm in depth is located at the center of the surface, and 120 output detectors with 200 nm $\times$ 200 nm in area and 40 nm in depth are regularly arranged along the $x$ and $y$ axes on the surface.  In numerical experiments with a micromagnetic simulator Mumax 3 \cite{vansteenkiste2014design}, the mesh was a cubic with $40 \times 40 \times 40\ \textup{nm}^3$ along the Cartesian coordinates, a spin having a saturation magnetization $M_\textup{S}$ was located at every mesh corner, the simulation time step was $1 \times 10^{-11}$ s, and temperature was set at 0 K.  We used material parameters consistent with those estimated from epitaxial YIG(111) films on GGG(111): $M_\textup{S} $ = 190 kA/m (at low temperatures \cite{anderson1964molecular}), the uniaxial anisotropy along the $z$ axis $K_\textup{U}^\textup{H}$ = 5.0 KJ/$\textup{m}^3$ \cite{manuilov2009pulsed}, the exchange stiffness constant $A_\textup{EX}$ = $3.7 \times 10^{-12}$ J/m \cite{klingler2014measurements}, and the cubic magneto-crystalline anisotropy $\sim$0. 
Hereafter, the $x$, $y$, and $z$ components of the normalized spin $s$ are expressed by $s_x$, $s_y$, and $s_z$, respectively.  
The YIG film was divided into two parts: the signal transmission and damping regions.  The damping region having a 40-nm width is located at the left and right sides (the boundaries at $x$ = $\pm 6 \ \mu$m) and the bottom plane (the boundary at $z$ = $-$320 nm).  The damping constant $\alpha$ was set at $\alpha_0$ = 0.001 in the transmission region, whereas it was varied from 0.001 to 1 in the damping region to change the spin wave dynamics.  An external magnetic field $H^\textup{EX}$ was applied along the $+y$ direction and its magnitude was constant during a simulation batch.  This study used $\mu_0 H^\textup{EX}$ = 2 $-$ 8 mT to explore various spin wave dynamics under smaller magnetic fields towards practical electronics. In the device operation, an input value was expressed by a corresponding $K_\textup{U}$ value in the exciter, while $K_\textup{U}^\textup{H}$ remains unchanged in the rest region. The output value at each detector was determined by the averaged $s_z$ inside. 
\vspace{-10pt}
\section{Characterization of spin dynamics}

\begin{figure}[b]
\includegraphics[width=8cm]{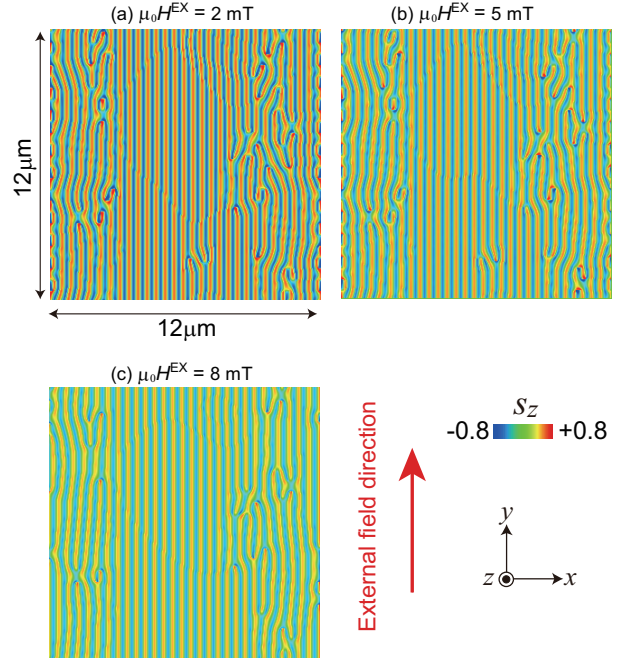}
\caption{\label{FIG_2} Top-view distributions of the $z$ component $s_z$ of the normalized spin $s$ in the magnetic garnet film under $\mu_0 H^\textup{EX}$ = (a) 2, (b) 5, and (c) 8 mT, respectively, where a Cartesian coordinate system is shown and a color bar represents the magnitude of $s_z$.}
\vspace{-10pt}
\end{figure}


 Figures \ref{FIG_2}(a), (b), and (c) show the top-view distributions of $s_z$ under $\mu_0 H^\textup{EX}$ = 2, 5, and 8 mT, respectively, where the Cartesian coordinate system is shown and the color bar represents the magnitude of $s_z$.  The stripe domain structures are basically similar to each other, whereas the absolute magnitude of $s_z$ decreases with the increase of $\mu_0 H^\textup{EX}$.  Our previous paper revealed that such domain structures are effective for RC \cite{nakane2021spin}.

\begin{figure}[tb]
\includegraphics[width=\linewidth]{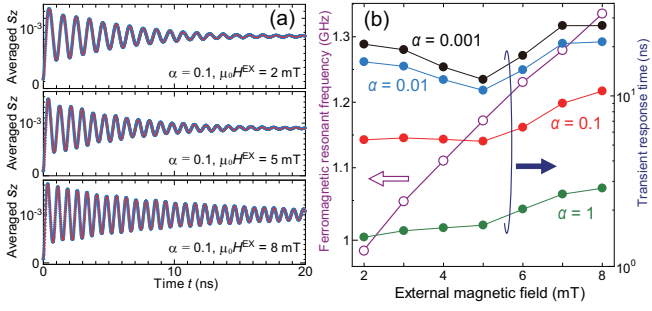}
\caption{\label{FIG_3} (a) Time evolution of the averaged $s_z$ after the application of a magnetic field along the $+z$ direction $\mu_0H^\textup{z}$ = 0.1 mT at $t$ = 0, where $\alpha$ = 0.1, $\mu_0H^\textup{EX}$ = 2, 5, and 8 mT (from top to bottom), blue dots represent calculated values, and a red curve at each $\mu_0H^\textup{EX}$ is the fitting curve. (b) Ferromagnetic resonant frequency $f_0$ and relaxation time $\tau_0$ plotted against $\mu_0 H^\textup{EX}$, where purple open circles represent $f_0$ values for all the alpha values and black, blue, red, and green closed circles represent the transient response time $\tau_0$ values for $\alpha$ = 0.001, 0.01, 0.1, and 1, respectively.}
\vspace{-10pt}
\end{figure}

Fundamental properties of spin dynamics were characterized by the ferromagnetic resonant (FMR) frequency $f_0$ and the relaxation time $\tau_0$ of the spin precession under $\mu_0 H^\textup{EX}$ (= 2 $-$ 8 mT in steps of 1 mT). After the relaxation of a magnetic state under a specific $\mu_0 H^\textup{EX}$ value, $\mu_0 H^\textup{z}$ = 0.1 mT along the $+z$ direction was added at $t$ = 0 and the averaged $s_z$ over the entire film was recorded at every 1 $\times 10^{-11}$ s.  Finally, $f_0$ and $\tau_0$ were estimated by fitting $A\textup{sin}(2\pi f_0 + \phi_0)\textup{exp}(-t/\tau_0)$ to the signal, where $A$ and $\phi_0$ are constants.
Figure \ref{FIG_3}(a) shows time evolution of the averaged $s_z$ after the application of $\mu_0H^\textup{z}$ = 0.1 mT at $t$ = 0, where $\alpha$ = 0.1, $\mu_0H^\textup{EX}$ = 2, 5, and 8 mT (from top to bottom), blue dots represent calculated values, and a red curve at each $\mu_0H^\textup{EX}$ is the fitting curve.  
Figure \ref{FIG_3}(b) shows $f_0$ and $\tau_0$ plotted against $\mu_0 H^\textup{EX}$, where purple open circles represent $f_0$ values for all the alpha values and black, blue, red, and green closed circles represent $\tau_0$ values for $\alpha$ = 0.001, 0.01, 0.1, and 1, respectively. Whereas $f_0$ monotonically increases with increasing $\mu_0 H^\textup{EX}$, $\tau_0$ varies with both $\mu_0 H^\textup{EX}$ and $\alpha$. This complicated properties are likely related with the stripe domain structure changing with $\mu_0 H^\textup{EX}$, as shown in Figs. \ref{FIG_2}(a), (b), and (c). Those $f_0$ and $\tau_0$ results indicate that the spin wave dynamics is changed with $\mu_0 H^\textup{EX}$ and $\alpha$. 

\begin{figure}[b]
\includegraphics[width=\linewidth]{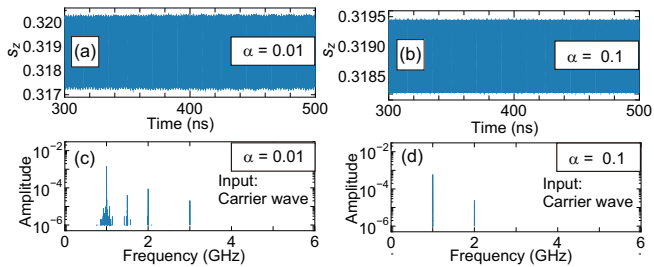}
\caption{\label{FIG_4} (a)(b) Output signals at a detector under $\mu_0 H^\textup{EX}$ = 5 mT and (c)(d) their discrete Fourier transform (DFT) spectra, when a time-series 1 GHz cosine wave (carrier wave) having an amplitude $\Delta K_\textup{U}$ = 500 $\textup{J/m}^3$ is fed into the input exciter: $\alpha$ = (a)(c) 0.01 and (b)(d) 0.1.}
\vspace{-10pt}
\end{figure}


Based on the $f_0$ results, a cosine wave with 1 GHz was used as a carrier wave for input signals.
To characterize fundamental properties of the carrier wave, 
a cosine wave with a constant amplitude $\Delta K_\textup{U}$ = 500 J/$\textup{m}^3$ was fed into the input exciter and output waveforms at various detectors were observed under $\alpha$ = 0.001, 0.01, 0.1, and 1. Figures \ref{FIG_4}(a) and (b) show output signals under $\mu_0 H^\textup{EX}$ = 5 mT and $\alpha$ = 0.001 and 0.1, respectively, which were obtained by \color{blue} a \color{black}detector at the center left in the right transmission area in Fig. \ref{FIG_1}. Figures \ref{FIG_4}(c) and (d) show discrete Fourier transform (DFT) spectra calculated from Figs. \ref{FIG_4}(a) and (b), respectively, for which the time length of an input signal was 1.3 $\mu$s, the output signals in the first 100 ns range were discarded, and the Hanning window was used. 
In all the $\alpha$ cases, nonlinear phenomena are characterized by components at integer hamonics in the DFT spectra. When $\alpha$ = 0.001 and 0.01, there are fluctuations in the amplitude of the waveforms, as seen in Fig. \ref{FIG_4}(a), which appear as spread of the components at 1GHz and interharmonics components in the DFT spectra in Fig. \ref{FIG_4}(c). 
These DFT features are likely related with RC performance, as will be discussed in Sec. IV-D. Note that the lower limit of the accuracy in DFT spectra was set at 1 $\times$ $10^{-6}$ in this study.

\section{Reservoir computing}
\subsection{Preprocessing and reservoir parts}
\begin{figure}[tb]
\includegraphics[width=\linewidth]{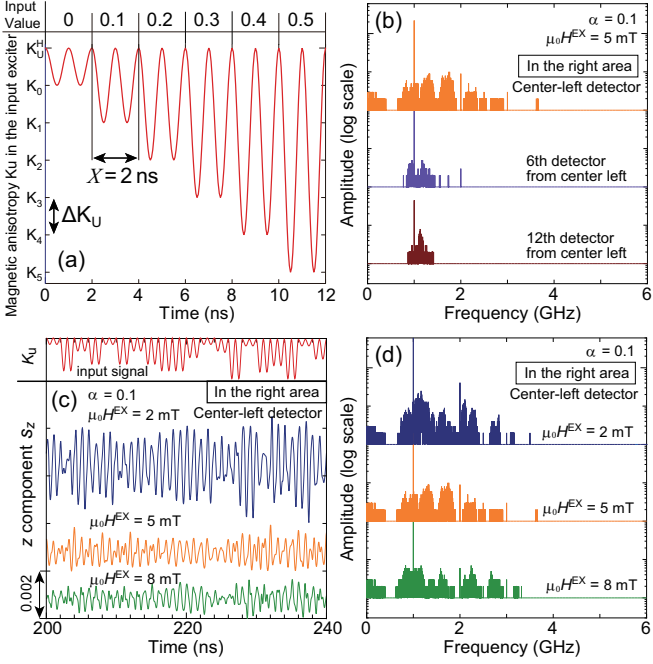}
\caption{\label{FIG_5} (a) Input value (= 0 $-$ 0.5) and the corresponding
input signal that is a time-series $K_\textup{U}$, where the carrier wave is a 1 GHz cosine wave, $X$ = 2 ns is the time step, the maximum $K_\textup{U}^\textup{H}$ = 5 $\textup{kJ/m}^3$, and $\Delta K_\textup{U}$ = 100 $\textup{J/m}^3$. (b) DFT spectra calculated from three output signals in the right transmission area (Fig. \ref{FIG_1}), where orange, purple, and brown curves were obtained by detectors at the center left, the 6th one from the center left, and the 12th one from the center left, respectively, $\mu_0 H^\textup{EX}$ = 5 mT, and $\alpha$ = 0.1. (c) Waveforms and (d) corresponding DFT spectra at the center-left detector in the right transmission area, where $\alpha$ = 0.1, dense-blue, orange, and green curves are the results under $\mu_0 H^\textup{EX}$ = 2, 5, and 8 mT, respectively, and a red curve in (c) represents the waveform of the input signal. In (b), (c), and (d), each curve is vertically shifted to facilitate visualization.}
\vspace{-10pt}
\end{figure}

In the preprocessing part, a sequential random input data $u(n)$ with a discrete time step $n$ (= 1, 2, $\dots$, 4100) was converted to a time-series input signal $K_\textup{U}(t)$ with a continuous time $t$: An input value (= 0, 0.1, 0.2, 0.3, 0.4 and 0.5) was converted to the corresponding $K_q  (q$ = 0, 1, 2, 3, 4, and 5) value with a constant interval $\Delta K_q$ = 100 J/$\textup{m}^3$.  
The amplitude of the carrier wave was modulated by $K_q$ with a time step length $X$ = 2 ns, as shown by a red curve in Fig. \ref{FIG_5}(a), where the amplitude in $(n-1)X \leq t < nX$ is $K_\textup{U}^\textup{H} - K_q(t)$.  In the reservoir part, $K_\textup{U}(t)$ was fed into the input exciter and the response output signal $x_i(t)$ at the $i$th detector ($i$ = 1, 2, $\dots$,120) was recorded at every $5 \times 10^{-11}$ s. 

In the reservoir part in Fig. \ref{FIG_1}, three output signals in the right transmission area are shown, where orange, purple, and brown curves were obtained by detectors at the center left, the 6th one from the center left, and the 12th one from the center left, respectively, $\mu_0 H^\textup{EX}$ = 5 mT, and $\alpha$ = 0.1. Those results confirm that various $x_i(t)$ waveforms were generated through spatially-distributed spin dynamics. 
Figure \ref{FIG_5}(b) shows DFT spectra calculated from the orange, purple, and brown signals in Fig. \ref{FIG_1}, for which the time length of an input signal was 8.2 $\mu$s, the output signals in the first 200 ns range were discarded, and the Hanning window was used.  Unlike the spectrum in Fig. \ref{FIG_4}(d), all the spectra in Fig. \ref{FIG_5}(b) have interharmonic components, despite of the same $\alpha$ value of 0.1. Thus, the information about the random 6-valued amplitude modulation in the input signal appears as interharmonic components in the output signals.

To analyze the effect of $\mu_0 H^\textup{EX}$ on output signals, waveforms at the center-left detector and their DFT spectra were observed for $\alpha$ = 0.1 and $\mu_0 H^\textup{EX}$ = 2, 5, and 8 mT, as shown in Fig. \ref{FIG_5}(c) and (d), respectively, where blue, orange, green curves represent the results for $\mu_0 H^\textup{EX}$ = 2, 5, and 8 mT, respectively, and a red curve in (c) represents the waveform of the input signal. Those results clearly characterize that the output signal at the same detector differs with $\mu_0 H^\textup{EX}$, whereas a peaked DFT component always appears at the forced oscillation frequency 1 GHz, irrespective of $f_0$ changing with $\mu_0 H^\textup{EX}$. Thus, it was found that the change in $\mu_0 H^\textup{EX}$ works well to efficiently generate different $x_i(t)$ at the single $i$th detector. Note that the amplitude of the waveform in Fig. \ref{FIG_5}(c) is the largest at $\mu_0 H^\textup{EX}$ = 2 mT, which is probably due to the fact that 1 GHz is nearly the same as the $f_0$ value for this condition (Fig. \ref{FIG_3}(b)).

It is noteworthy that this study uses a 1 GHz cosine wave as a carrier wave instead of time-series triangular pulses in our previous paper\cite{nakane2021spin}. From the analyses in both cases, harmonic components are not necessary in an input signal to realize diverse nonlinear dynamics in output signals. Besides, nonlinear phenomena were also obtained even when $f_0$ significantly differs from the input carrier frequency ($\mu_0 H^\textup{EX}$ = 5 and 8 mT conditions). Thus, it is expected that a wide range of carrier wave parameters can be used. This scalability can be a technical advantage of the spin-wave-based device. 


\subsection{Readout part}
In the readout part, $x_i(t)$ were converted to the time-step reservoir output state $y_i(n)$ ($n$ = 1, 2, $\dots$, 4100) using a postprocessing step: an envelope extraction, a low-pass filtering, and the averaging signal over $X$.  Hereafter, $x_i(t)$ and $y_i(n)$ obtained under $\mu_0 H^\textup{EX}$ = 2, 5, and 8 mT are denoted by $x_{j, i}(t)$ and $y_{j, i}(n)$, corresponding to $j$ = 1, 2, and 3, respectively.  To exclude unstable responses, $y_{j, i}(n)$ in the first 100 steps were discarded, $y_{j, i}(n)$ in $101 \leq n \leq 3598$ were used for the training, and the remaining $y_{j, i}(n)$ in $3599 \leq n \leq 4098$ were used for the testing.  Note that $y_{j, i}(n)$ in the last two $n$ steps were also discarded because of higher error in the envelope processing.
Next, the reservoir output vector $\mathbf{y}_j(n)$ was given through a row-wise collection of $y_{j,i}(n)$:
\begin{eqnarray}
\mathbf{y}_j(n) &=& (y_{j,1}(n), \dots, y_{j, 120}(n))^\mathsf{T} \in \mathbb{R}^{120}.
\end{eqnarray}
Then, combination of $\mathbf{y}_j(n)$ with different $j$ values were considered as follows:
$$ \mathbf{y}(n) = 
    \begin{cases}
        {\mathbf{y}_1(n),  \mathbf{y}_2(n),  \textup{or} \ \mathbf{y}_3(n) \hspace{20mm} \textup{for} \ N_\textup{y} = 120} \\
         {[\mathbf{y}_1(n); \mathbf{y}_2(n)],  [\mathbf{y}_2(n);  \mathbf{y}_3(n)],  \textup{or} \ [\mathbf{y}_3(n); \mathbf{y}_1(n)]} \\
         {\hspace{49.5mm} \textup{for} \ N_\textup{y} = 240} \ ,\\
        {[\mathbf{y}_1(n); \mathbf{y}_2(n); \mathbf{y}_3(n)] \hspace{22mm} \textup{for} \ N_\textup{y} = 360}
    \end{cases}
$$
where $N_\textup{y}$ is the dimension of a vector and $[\mathbf{a};\mathbf{b}]$ represents a vertical concatenation of vectors.
Using the system output at the $l$th node $z_l(n)$, a system output vector $\mathbf{z}(n)$ was given by:
\begin{eqnarray}
\mathbf{z}(n) &=& (z_1(n), \dots, z_{N_z}(n))^\mathsf{T} \in \mathbb{R}^{N_z}.
\end{eqnarray}
The output weight matrix $\mathbf{W}^\textup{out}$ was defined as follows:
\begin{eqnarray}
\mathbf{W}^\textup{out} &=& (w^\textup{out}_{li}) \in \mathbb{R}^{N_z \times N_y}.
\end{eqnarray}
Then, the system output was given by:
\begin{eqnarray}
\mathbf{z}(n) &=& \mathbf{W}^\textup{out}\mathbf{y}(n).
\end{eqnarray}
In the training phase, the target of the system output vector $\mathbf{d}(n)$ was given by a collection of the target (desired) signal $d_l(n)$ for the $l$th system output:
\begin{eqnarray}
\mathbf{d}(n) &=& (d_1(n), \dots, d_{N_z}(n))^\mathsf{T} \in \mathbb{R}^{N_z}.
\end{eqnarray}
The target matrix $\mathbf{D}$ and reservoir output matrix $\mathbf{Y}$ were formed by column-wise collections of $\mathbf{d}(n)$ and $\mathbf{y}(n)$ for $101 \leq n \leq 3598$, respectively.  The optimum weight matrix $\hat{\mathbf{W}}^{\textup{out}}$ was calculated by \cite{lukovsevivcius2009reservoir}:
\begin{eqnarray}
\hat{\mathbf{W}}^{\textup{out}} &=& \mathbf{D}\mathbf{Y}^{\dagger},
\end{eqnarray}
where $\mathbf{Y}^{\dagger}$ is the Moore-Penrose pseudoinverse of $\mathbf{Y}$.  In the testing phase, the system output vector $\mathbf{z}^*(n)$ was obtained by:
\begin{eqnarray}
\mathbf{z}^*(n) &=& \hat{\mathbf{W}}^{\textup{out}}\mathbf{y}(n).
\end{eqnarray}

\subsection{Benchmark tasks}
We performed two time-series benchmark tasks that are widely used for evaluating RC systems.  The first one is to predict  time series data generated from the $m$th-order nonlinear autoregressive moving average (NARMA-$m$) model ($m = 2, \dots, 10$) derived from Eq. (86) in ref. \cite{atiya2000new}.  A time-series target $d_m(n+1)$ = $z_m(n+1)$ was iteratively generated using the following relation:
\begin{multline}
z_m(n+1) = 0.3z_m(n) + 0.5z_m(n)\sum_{s = 0}^{m-1}z_m(n-s) \\
+ 1.5u(n-m-1)u(n) +1.
\end{multline}
The performance was evaluated by the normalized mean square error ($\textup{NMSE}_m$) for the NARMA-$m$ task:
\begin{eqnarray}
\textup{NMSE}_m &=& \frac{1}{L} \frac{\sum_{n = 1}^{L}(d_m(n) - z_m^*(n))^2}{\textup{Var}(d_m(n))}, 
\end{eqnarray}
where Var denotes variance and $L$ = 500 is the step length in the testing duration.
The second one is delay tasks whose targets are given by:
\begin{eqnarray}
d_k(n) = u(n-k) \hspace{0.5cm} \mbox{for delay} \, k = 1,\ldots, 120.
\end{eqnarray}
The performance was evaluated by memory capacity \cite{jaeger2002short}:
\begin{eqnarray}
\textup{MC}_k &=& \frac{\textup{Cov}^2(u(n-k), z^*(n, k))}{\textup{Var}(u(n))\textup{Var}(z^*(n, k))}, 
\end{eqnarray}
where Cov and $z^*(n, k)$ denote covariance and the predicted value for $d_k(n)$, respectively, and the time step length was 400.  The total MC is the summation of $\textup{MC}_k$ over $k$ = 1 $-$ 120. 


\subsection{Results and analysis}
In Fig. \ref{FIG_1}, time-evolutional NARMA-10 signals are shown in the readout part, where gray and blue curves represent the target $d_{10}(n)$ and predicted $z^*_{10}(n)$ signals in the testing phase, respectively, under $N_\textup{y}$ = 360 and $\alpha$ = 0.1. In the figure, $z^*_{10}(n)$ reasonably reproduces $d_{10}(n)$. Hence, many $x_{j, i}(t)$ have both short-term memory and nonlinearity effective for RC and the post signal processing extracted such properties for high dimensionality in $\mathbf{y}(n)$.


\begin{figure}[b]
\includegraphics[width=\linewidth]{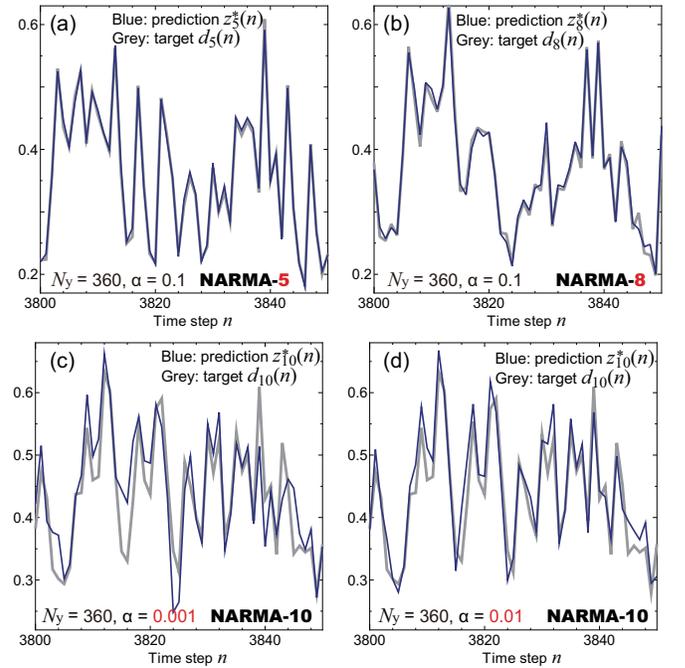}
\caption{\label{FIG_6} (a) NARMA-5 and (b) NARMA-8 signals under $N_\textup{y}$ = 360 and $\alpha$ = 0.1. (c)(d) NARMA-10 signals under  $N_\textup{y}$ = 360 and $\alpha$ = (c) 0.001 and (d) 0.01. The gray and blue curves represent the target $d_m(n)$ and prediction $z^*_m(n)$ signals, respectively.}
\end{figure}

\begin{figure}[tb]
\includegraphics[width=\linewidth]{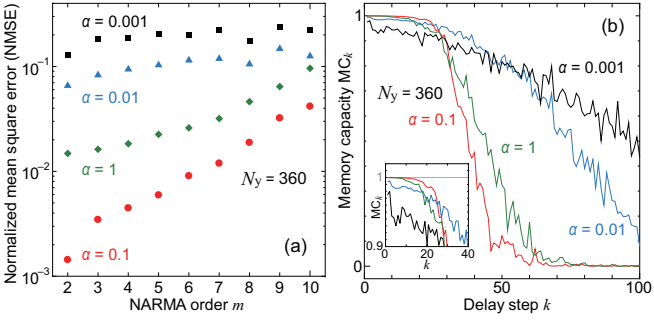}
\caption{\label{FIG_7} (a) Normalized mean square error for $m$th-order NARMA ($\textup{NMSE}_m$) plotted against $m$ and (b) memory capacity $\textup{MC}_k$ plotted against time-step delay $k$ when $N_\textup{y}$ = 360, where black, blue, red, and green colors are estimated values for $\alpha$ = 0.001, 0.01, 0.1, and 1, respectively. The inset is a close-up view of $\textup{MC}_k$ $\sim$1 at around $k$ = 0.}
\vspace{-10pt}
\end{figure}

Figures \ref{FIG_6}(a) and (b) show NARMA-5 and NARMA-8 signals under the same $N_\textup{y}$ = 360 and $\alpha$ = 0.1 condition, respectively, where gray and blue curves represent the target $d_m(n)$ and prediction $z^*_m(n)$ signals, respectively. For lower $m$ values, $z^*_m(n)$ more accurately reproduces $d_m(n)$. This is reasonable since the prediction becomes harder for the task with higher $m$. Figures \ref{FIG_6}(c) and (d) show NARMA-10 signals under $N_\textup{y}$ = 360 and $\alpha$ = 0.001 and 0.01, respectively. It was found that $z^*_m(n)$ more accurately reproduces $d_m(n)$ with the increase of $\alpha$ from 0.001 to 0.1.
In the same manner, NARMA-$m$ signals were calculated for $N_\textup{y}$ = 360 and $\textup{NMSE}_m$ were plotted against $m$, as shown in Fig. \ref{FIG_7}(a), where black squares, blue triangles, red circles, and green diamonds are estimated values for $\alpha$ = 0.001, 0.01, 0.1, and 1, respectively.  In the entire $m$ range, $\textup{NMSE}_m$ decreases as $\alpha$ is increased from 0.001 to 0.1, whereas it increases as $\alpha$ is increased from 0.1 to 1. 

 To analyze the results in Fig. \ref{FIG_7}(a), $\textup{MC}_k$ was plotted against $k$, as shown in Fig. \ref{FIG_7}(b), where black, blue, red, and green curves are estimated memory curves for $\alpha$ = 0.001, 0.01, 0.1, and 1, respectively. As $\alpha$ is decreased from 0.1 to 0.001, the $k$ length for $\textup{MC}_k$ $\sim$1 decreases and the slope of the curve becomes more gradual. This trend is very similar to that of the memory curve with increasing noise in an input sequence in ref. \cite{jaeger2002short}. This can be interpreted as follows: The fluctuation of the carrier wave increases with the decrease of $\alpha$ from 0.01 to 0.001 and it acts as noise in the derivation of $\hat{\mathbf{W}}^{\textup{out}}$ because the fluctuation and the random 6-valued input information are characterized by interharmonic components in Figs. \ref{FIG_4}(c) and \ref{FIG_5}(b), respectively. This $``$noise effect$"$ is probably related with the high $\textup{NMSE}_m$ values.


On the other hand, the difference between the results with $\alpha$ = 0.1 and 1 possibly originates from the difference in $\tau_0$: $\tau_0 = 5.3 - 10.7$ ns for $\alpha$ = 0.1 and $\tau_0 = 1.5 - 2.7$ ns for $\alpha$ = 1 (Fig. \ref{FIG_3}(b)). It is expected that the response signals were significantly relaxed during  $X$ = 2 ns when $\alpha$ = 1.  For RC, nonlinear responses with short-term memories of past inputs are necessary.  Thus, although it has not been fully clarified, many studies revealed that the time step $X$ moderately smaller than $\tau_0$ is important for high computing capabilities \cite{tanaka2019recent, appeltant2011information, torrejon2017neuromorphic}. In this respect, the $\alpha$ = 1 condition does not meet the requirement. We consider that the highest performance with $\alpha$ = 0.1 was achieved through the above two beneficial conditions: no interharmonics components in the carrier wave and $X < \tau_0$.

The relation between $\textup{NMSE}_m$ in Fig. \ref{FIG_7}(a) and $\textup{MC}_k$ in Fig. \ref{FIG_7}(b) was further analyzed.  
When the total MC for the $\alpha$ values are compared with each other, the $\alpha$ = 0.1 condition has the smallest value of 38.  However, when $\textup{MC}_k$ at around smaller $k$ is focused, as shown in the inset, the $\alpha$ = 0.1 condition has the longest $k$ length for $\textup{MC}_k$ $\sim$1.  Hence, the highest performance in the NARMA tasks for $\alpha$ = 0.1 can be also characterized by the highly accurate memory for the longest $k$ length in $\textup{MC}_k$.  From the results thus far, the $\alpha$ = 0.1 condition is focused below.


\begin{figure}[b]
\includegraphics[width=\linewidth]{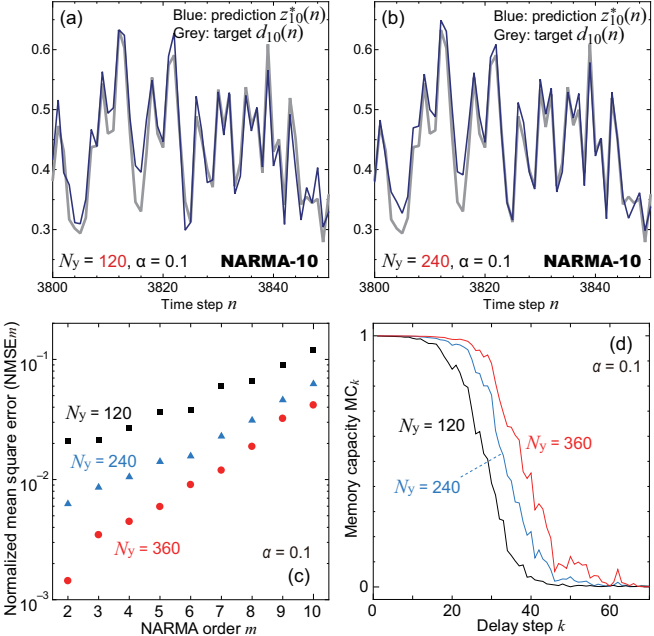}
\caption{\label{FIG_8} (a)(b) NARMA-10 signals under $\alpha$ = 0.1 and $N_\textup{y}$ = (a) 120 and (b) 240, where gray and blue curves represent the target $d_{10}(n)$ and prediction $z^*_{10}(n)$ signals, respectively. (c) $\textup{NMSE}_m$ plotted against $m$ and (d) $\textup{MC}_k$ plotted against $k$, when $\alpha$ = 0.1, where black blue, and red colors represent the results estimated for $N_\textup{y}$ = 120, 240, and 360, respectively.}
\end{figure}


Figures \ref{FIG_8}(a) and (b) show NARMA-10 signals under $\alpha$ = 0.1 and $N_\textup{y}$ = 120 and 240, respectively, where gray and blue curves represent the target $d_{10}(n)$ and prediction $z^*_{10}(n)$ signals, respectively. Those two figures confirm that $z^*_{10}(n)$ reasonably reproduces $d_{10}(n)$ even under $N_\textup{y}$ = 120 and the difference in the computational performance between $N_\textup{y}$ = 120 and 240 is seemingly insignificant. 
On the other hand, there are three different $\textup{NMSE}_m$ and $\textup{MC}_k$ values when $N_\textup{y}$ = 120 and 240.  In each case, the discrepancy between the results were within the error margin of $\sim$20$\%$.
Hereafter, the averaged result will be shown.

Figure \ref{FIG_8}(c) shows $\textup{NMSE}_m$ plotted against $m$, where black squares, blue triangles, and red circles represent the values for $N_\textup{y}$ = 120, 240, and 360, respectively.  In the entire $m$ range, $\textup{NMSE}_m$ decreases as $N_\textup{y}$ is increased, indicating that the dimensionality effective for NARMA-$m$ tasks increases with $N_\textup{y}$.  The effect of $N_\textup{y}$ on $\textup{MC}_k$ was also analyzed by $\textup{MC}_k$$- \ k$ plot, as shown in Fig. \ref{FIG_8}(d), where black, blue, and red curves are estimated memory curves for $N_\textup{y}$ = 120, 240, and 360, respectively.  It was found that both the total MC and the $k$ length for $\textup{MC}_k$ $\sim$1 increase with increasing $N_\textup{y}$.  Thus, $N_\textup{y}$ is probably the dominant parameter for high computational performance in various computational tasks.  
\vspace{-10pt}

\section{Discussion}
At $m$ = 10 (NARMA-10), the $\textup{NMSE}_{10}$ values for $N_\textup{y}$ = 120, 240, and 360 are 0.120, 0.063, and 0.042, respectively, when $\alpha$ = 0.1. The minimum $\textup{NMSE}_{10}$ = 0.042 is comparable to those obtained in the most advanced physical RC that utilizes optical systems with a feedback loop \cite{chembo2020machine}. To the best of our knowledge, NARMA-10 prediction has not been demonstrated by an on-chip RC device, even in numerical experiments (excluding an electronic circuit simulation \cite{appeltant2011information}). To further gain insight into the performance enhancement with increasing $N_\textup{y}$, the $\textup{NMSE}_{10}$ values are discussed through the comparison with those estimated using echo state networks (ESNs) with some network sizes in ref. \cite{rodan2010minimum} : $\textup{NMSE}_{10}$  values are 0.095, 0.051, and 0.042 for $N_\textup{y}$ = 100, 150, and 200, respectively.  Although our $\textup{NMSE}_{10}$ value at similar $N_\textup{y}$ values is a little larger than that for the ESN, $\textup{NMSE}_{10}$ tends to saturate with increasing $N_\textup{y}$ in both cases.  Thus, it is a general property that the enhancement rate decreases with increasing $N_\textup{y}$. 

\begin{figure}[tb]
\includegraphics[width=\linewidth]{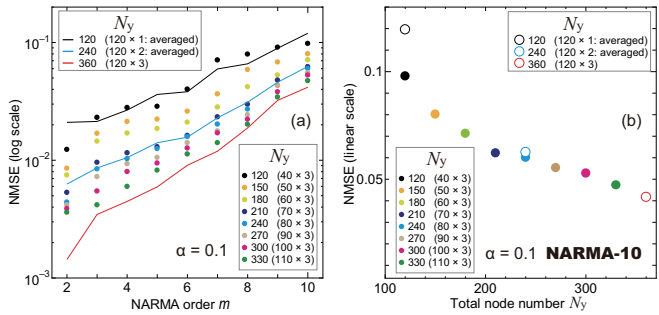}
\caption{\label{FIG_9} (a) $\textup{NMSE}_m$ plotted against $m$ when $\alpha$ = 0.1, where colored circles are the results for $\mathbf{y}(n)$ having $y_{1,i}(n)$, $y_{2,i}(n)$, and $y_{3,i}(n)$ ($i \leq N_\textup{y}/3$).  The black-, blue-, and red-solid lines are the results for $\mathbf{y}(n)$ with $N_\textup{y}$ = 120, 240, and 360 in Fig. \ref{FIG_8}(c), respectively. (b) $\textup{NMSE}_{10}$ plotted against $N_\textup{y}$, where colored circles correspond to those in (a) and open circles correspond to those in Fig. \ref{FIG_8}(c).}
\end{figure}

For further study, the vectors $\mathbf{y}(n)$ with various $N_\textup{y}$ (= 120 $-$ 330) values were constructed by collecting $y_{j,i}(n)$ with $j$ (= 1, 2, 3) and $i$ (= 1, 2, $\dots$ $N_\textup{y}/3$). Figure \ref{FIG_9}(a) shows $\textup{NMSE}_m$ plotted against $m$, where colored circles are the results for $\mathbf{y}(n)$ having $y_{1,i}(n)$, $y_{2,i}(n)$, and $y_{3,i}(n)$ ($i \leq N_\textup{y}/3$).  The results for $\mathbf{y}(n)$ with $N_\textup{y}$ = 120, 240, and 360 in Fig. \ref{FIG_8}(c) are also plotted by black-, blue-, and red-solid lines, respectively.  With the increase of $N_\textup{y}$, $\textup{NMSE}_m$ decreases at all the $m$ values. At $N_\textup{y}$ = 120 and 240, the circles and lines are remarkably similar to each other regardless of the selection rule, indicating that each $y_{j,i}(n)$ plays a comparable role in computing. Figure \ref{FIG_9}(b) shows $\textup{NMSE}_{10}$ plotted against $N_\textup{y}$, where colored circles correspond to those in Fig. \ref{FIG_9}(a) and open circles correspond to those in Fig. \ref{FIG_8}(c). As $N_\textup{y}$ is increased, $\textup{NMSE}_{10}$ continuously decreases and its rate of change seemingly decreases. 
The features in Figs. \ref{FIG_9}(a) and (b) clearly reveal that every $y_{j,i}(n)$ contributes to the computational performance, since $N_\textup{y}$ simply determines $\textup{NMSE}_m$.  Therefore, it can be concluded that the performance enhancement in our system originates from the increase in the dimensionality in $\mathbf{y}(n)$.  

Considering that our method employs the spin dynamics under the different conditions in parallel, the whole system can be interpreted as a multi-reservoir system such as the grouped echo state networks \cite{gallicchio2017deep, li2022multi}. The explanation in terms of system architecture is our future interest.

\section{conclusion}



We have numerically studied the performance enhancement in RC utilizing the single spin-wave-based reservoir device with the 1-input exciter and 120-output detectors. The input signal was the 1 GHz cosine wave whose amplitude was modulated following a series random 6 values and the output signals at the detectors were found to have nonlinear phenomena changing with the spatial position, external magnetic field, and  damping constant at the sides and bottom of the magnetic garnet film. The reservoir output vectors with various dimensions, including higher than 120 dimension, were constructed by collecting the output signals under the three different magnetic fields. With the optimum damping constant, the computational performance in NARMA tasks was enhanced with the increase in the dimension of the reservoir output vector. The result is evidence that our method efficiently increases the dimensionality effective for NARMA tasks.

We demonstrated that the single spin-wave-based device with the $\mu_0 \mathbf{H}^{\textup{EX}}$-controlled different physical conditions can generate diverse output signals that are essential for higher computational performance. This strategy is also very useful for other physical RC, along with the virtual node method, to tackle nonlinear time series prediction tasks. In particular, since it is quite strict to implement a vast number of real output nodes on a chip device, performance enhancement through various configuration settings is a practical approach.


\begin{acknowledgments}
This work was supported in part by the New Energy and Industrial Technology Development Organization (NEDO) under Project JPNP16007.
\end{acknowledgments}
\bibliography{aipsamp}

\end{document}